\documentclass[12pt]{article}
\usepackage{epsf}
\usepackage{slashed}
\usepackage{xcolor}
\usepackage{hyperref}
\begin{document}
\begin{titlepage}
\title{ Modified Newtonian dynamics and hadronic scales}
\author{
{Piotr \.Zenczykowski }\footnote{E-mail: piotr.zenczykowski@ifj.edu.pl}\\
{\em Professor Emeritus}\\
{\em The Henryk Niewodnicza\'nski Institute of Nuclear Physics}\\
{\em Polish Academy of Sciences}\\
{\em Radzikowskiego 152,
31-342 Krak\'ow, Poland}\\
}
\maketitle
\begin{abstract}
In this paper, we use the idea of modified Newtonian dynamics to further support
the arguments that important information concerning the nature of space
is hidden at hadronic mass and distance scales.\\
\end{abstract}

\vfill
\end{titlepage}

In paper \cite{Zen2018}  the issue of a possible 
connection between the hadronic world and the idea of space
emergence was discussed at some length. As the conceptual starting point we adopted the Aristotelian position  which assumes  logical priority of matter over space. In other words it was accepted that space should be viewed as an attribute of matter, a consequence of its existence. 
We argued in favour of this position by 
pointing out that in the past it was assumed by many philosophically-minded thinkers such as Leibniz, Mach, Einstein,\footnote{As Einstein put it for the general public in 1921:
{\it ``time and space disappear together with things''} (see, eg. \cite{Einstein}).} Heisenberg, and others, contributing in particular to the development of general relativity (GR).
According to this idea, properties of space should be connected with and derivable from those of matter (see also eg. \cite{Pen1968}).\\ 

The quantum properties of matter and the discretization of elementary particle masses 
as well as the widespread belief in the general applicability of quantum ideas `in the small'
suggest then some form of space quantization (or discretization) at `sufficiently small' distances.
In other words, it is often conjectured that at some fundamental quantum level there is no continuous spacetime that provides a background for the more familiar physical processes, 
and that this background emerges only in
some (not yet defined) high complexity limit, and in particular at appropriately large distances.
Lacking an experimental input to guide a more precise development of this vague conception,
theoretical speculations on the idea of space emergence range then over a perplexingly wide variety of approaches, such as Penrose's twistors, string theory, loop quantum gravity, causal set theory, and many, many others. 
 Dimensional considerations
that accept the relevance of quantum ($h$) and spacetime (relativistic ($c$) and gravitational ($G$)) constants seem to indicate that the transition from the quantum to the classical description of space should occur at the Planck length $l_P=\sqrt{hG/c^3}=4.05 \times 10^{-33}~cm$, which is often considered as a scale relevant for space alone (ie. independent from matter/particle scales). \\

As an Aristotelian alternative to this view,
in \cite{Zen2018} several arguments were presented to the point that the relevant distance and mass scales may equally well be those appropriate for the description of hadrons, quarks and other elementary particles. In fact, such an opinion was expressed much earlier by Penrose who
said in \cite{Pen1974}: 
{\it ``(...) it seems likely that if twistors do turn out to provide a better formalism for
microphysics than does the conventional space-time approach, then we shall begin to
see the effects of this on the much larger scale of elementary particles (say $10^{-13}~cm)$"}.
In \cite{Pen1974} Penrose dismissed also the arguments that the hadronic distance scale is too large: {\it ``It has been argued 
that the agreement between quantum electrodynamics and experiment shows that the normal
space-time description of nature must hold true down below $10^{-15}$~cm, but this should
not be regarded as in any way contradicting the above statement. (...) It is the proton
itself, not the spacetime point, which behaves as a discrete physical entity and which has,
at least to a considerable degree, some semblance of indivisibility."}
Accepting the primary role of matter (ie. `proton'),
Penrose turned away from the Democritean spacetime-background description of nature in favour of a more Aristotelian approach.
He thought that there should exist a description of reality which is conceptually deeper
than the currently accepted (and very successful) Standard Model (SM) of elementary particles
of which quantum electrodynamics is just a part.

The arguments used in \cite{Zen2018} involved the consideration of other constants  in addition to $G,h$, and $c$
(such as the cosmological constant $\Lambda$  and the Regge hadronic slope $\alpha'$)
\footnote{Possible relevance of constants other than $G$, $c$, and $h$ was pointed out by Meschini in \cite{Meschini}.}, the generalization of the concept of mass as suggested by the phase-space-based explanation 
\cite{Zen2008,ZenBook,Clifford}
of the Harari-Shupe rishon model \cite{HSmodel} of elementary particles, and the phenomenological conclusions on the internal spatial structure of excited baryons
\cite{Capstick}.
In this note we want 
to shed some additional light on the argument involving the cosmological constant $\Lambda$.
\\

We know that  matter and its
mutual interactions are
described in the modern Democritean approach with the help of fields defined on the underlying spacetime - both `in the large' 
 and `in the small' (ie. in the SM). 
If Aristotelian spirit is to be invoked, the relevant background space should be regarded 
as an attribute (or product) of matter.
In order to preclude the classically unwanted action-at-a-distance,
the influence of matter on the existence and properties of the surrounding space is expected to proceed through 
the intermediary of the field it generates. The gravitational field should be most relevant here as only this field interacts with all kinds of matter.\\

Since all our physical concepts should be considered adequate only within some limited domains,
we may expect that the standard conception of space 
should become inapplicable for distances that are either sufficiently small or sufficiently large in comparison to typical macroscopic scales. Here we may tentatively think of two limits:
in `the large' - the observed size of the Universe ($l_U = 10^{29}cm$), and in `the small' - the  Planck length $l_P$.
In a more refined Aristotelian view, space is
a consequence of the existence of matter and the gravitational field it generates. 
Accordingly, the upper and lower limits 
beyond which our standard 
conception of space
is expected to be no longer applicable 
should not concern directly the distance scales but rather the scales of 
particle masses or the strength of gravitational field (or both). 
Since the strength of the gravitational field is described by the acceleration it induces, 
the relevant limits 
should probably be expressed as lower and upper limits
on the acceleration itself (from which conditions the limits on the relevant distances should follow).
As these limits one may tentatively choose:
\begin{equation}
a_P= c^2/l_P= 2.25 \times 10^{53} cm/s^2,
\end{equation} 
 and
\begin{equation}
\label{aU}
a_U =c^2/l_U = 10^{-8} cm/s^2.
\end{equation}
\\

Now, while we have no experimental means to learn what really happens at the distances and accelerations of the order
of $l_P$ and $a_P$, we do have satisfactory access to the limit of very weak accelerations.
This access is provided by astrophysical observations which suggest that something peculiar does happen at 
\begin{equation}
\label{aM}
a_M=1.2 \times 10^{-8}cm/s^2. 
\end{equation}
Specifically, it turns out that many astrophysical data can be extremely well
described by a simple MOdification of Newtonian Dynamics (MOND), an idea put forward almost 40 years ago by Milgrom
\cite{Milgrom1983}. The original proposal was triggered by the unexpected shape of stellar rotation curves that describe the behaviour of stars in the outer reaches of galactic structures. Namely, the data indicate that 
these stars move too fast when compared with the expectations based on the assumption that
the relevant gravitational forces are Newtonian in character and due solely to the observed material content of the galaxies in question. Thus either galaxies should be viewed as placed in huge halos of undetected 
(and therefore non-luminous and interacting extremely weakly)
`dark matter', or the gravitational forces (accelerations) at the relevant distances are significantly stronger 
than the Newtonian ones. 
Detailed analyses indicate that for a range of reasons 
the second  option (MOND) may be regarded as a strong competitor of the currently popular dark matter paradigm \cite{McGaugh2014}. In fact, MOND is being considered by its proponents  as  vastly superior to the dark matter idea.
For details and the presentation of a long list of relevant pro-MOND arguments see eg. \cite{McGaugh}. {To this  list one should add the most recent detection of the external field effect in galactic structures \cite{EFE2020}. The relevant observation supplied such arguments with a substantial boost
as the effect in question is very specific to MOND and does not appear in Newton-Einstein gravity.}\\

According to the proposed
modification of Newtonian dynamics, below the critical value of $a_M$ given in Eq. (\ref{aM}) the acceleration of a test body placed in the gravitational field of mass $m$
ceases to be given by the Newtonian expression $a=G m/r^2 \equiv a_N$.  
Instead it becomes \footnote{In the vicinity of $a=a_M$ there should be a small region of smooth transition between the Newtonian form $a_N$ and the `deep-MOND' form of Eq. (\ref{MOND1}).}
\begin{equation}
\label{MOND1}
a = \frac{\sqrt{Gm}}{r}\sqrt{a_M}.
\end{equation}
Comparing Eq. (\ref{MOND1}) with the expression for $a_N$ we see that for an object of mass $m$ this transition is centered
around
\begin{equation}
\label{rM}
r_M(m) = \sqrt{\frac{Gm}{a_M}}.
\end{equation}
Note that the distance $r_M(m)$ at which the realm of Newtonian dynamics is supposed to give way to the deep-MOND
region depends on the size of the field-generating mass $m$. In other words the transition between the two realms does {\it not} occur at some source-mass-independent universal distance. Instead, it occurs for the {\it extremely weak field} as specified by universal acceleration $a_M$.

One may wonder what happens when one goes from the astrophysical regime to the region of smaller and smaller masses $m$ (and the induced distances $r_M(m)$). Assuming that formula (\ref{rM}) is not specific for the mass and distance scales encountered in astrophysics but is quite {\it universal} and holds
over a very wide range of masses and distances (a crucial assumption in the following discussion) one obtains that in the Solar System (ie. with Sun as the source mass) the effects of MOND might appear for distances $r > r_M({m_\odot}) = 0.1$ light years only (ie. in the Oort cloud). 
For source mass of the order of $m_X=200~mg$ (as in \cite{NewtonExp}) the deep-MOND
region of sufficiently weak fields 
occurs for $r > r_M(m_X) \approx 1~cm$ (from (\ref{rM})).\footnote{ This is not the region accessed in experiments \cite{NewtonExp} in which the Newtonian $1/r^2$ law was confirmed at small distances ($50 \mu m$): at this distance the gravitational field of mass $m_X$ is still substantial (and the induced acceleration much larger than $a_M$).} In other words,
thinking about the change of the form of gravitational forces at some source-mass-independent and well-defined small distance may be highly misleading.\\

Now, the dominant view is that 
for sufficiently small distances we should ultimately reach the realm of the quantum description.
In the quantum description the characteristic distance associated with mass $m$ is
given by the relevant Compton wavelength:
\begin{equation}
\label{Compton}
r_C(m)=\frac{h}{mc}
\end{equation}
(which formula is accepted to summarize the quantum properties of matter over a very
wide range of masses and distances).\\

The transition between the Newtonian and the deep-MOND regions occurs at the characteristic
for the quantum description distance of $r_C(m)$ if 
\begin{equation}
\label{CM}
r_C(m)=r_M(m)\equiv r_{CM}.
\end{equation} 
From (\ref{CM}) one finds
the relevant mass and distance:
 \begin{equation}
\label{MCM}
m_{CM}=\left(\frac{h^2a_M}{Gc^2}\right)^{1/3} = 0.2 \times 10^{-24}g \approx m_H,
\end{equation}
\begin{equation}
\label{rCM}
r_{CM}=\frac{h}{m_{CM}c}=\left(\frac{Gh}{a_Mc}\right)^{1/3}=10^{-12} cm \approx r_H.
\end{equation}
One notes that $m_{CM}$ is of the order of a typical hadronic mass $m_H$
(the nucleon mass is $m_n=1.67\times 10^{-24}g$, the pion mass is $m_{\pi}=0.25\times 10^{-24}g$).
Likewise, $r_{CM}$ is of  the order of  the range $r_H$ of hadronic forces.\\ 

If instead of (\ref{rM}) (that corresponds to the transition from the Newtonian region to the weak-field deep-MOND regime) one 
takes the Schwarzschild radius 
\begin{equation}
\label{rS}
r_S(m) \approx Gm/c^2
\end{equation}
(relevant for the transition to the strong-field black-hole regime), and subsequently considers the counterpart of (\ref{CM}), ie.
\begin{equation}
\label{CS}
r_C(m)=r_S(m)\equiv r_{CS},
\end{equation}
  one finds - in place of (\ref{MCM}) and (\ref{rCM}) - that
\begin{equation}
\label{MCS}
m_{CS}=m_P=\sqrt{hc/G}=5.46 \times 10^{-5}g, 
\end{equation}
and 
\begin{equation}
\label{rCS}
r_{CS}=l_P= c^2/a_P=\sqrt{Gh/c^3}=\sqrt{G (hc)^{-1}(h/c)^2}=4.05 \times 10^{-33}~cm.
\end{equation}

{For illuminative reasons conditions (\ref{CM}) and (\ref{CS}) should be considered in connection with the
third condition of this type, namely with
\begin{equation}
\label{MS}
r_{M}(m)=r_{S}(m)\equiv r_{MS}.
\end{equation}
One finds that
\begin{equation}
\label {rMS}
r_{MS} = c^2/a_M = 7.5 \times 10^{28}~cm \approx r_U,
\end{equation}
and 
\begin{equation}
\label{mMS}
m_{MS}=\frac{c^4}{Ga_M} \approx 10^{56}~g \approx m_U,
\end{equation}
where $r_U$ and $m_U$ denote the size and mass of the Universe.\\

The three points, namely (1) Planck: $(r_{CS},m_{CS})$, (2) Universe: $(r_{MS},m_{MS})$, and (3) Hadron:
$(r_{CM},m_{CM})$ define the vertices of a triangle on the $(log~r, log~m)$ phase space plot that was considered for purely astrophysical reasons in \cite{Hernandez}. (In \cite{Hernandez}
the relevant plot
involved only the vertex $(r_{MS},m_{MS})$ and the associated angle.)  The interior of this `Newtonian' triangle\footnote{except for the vicinity of its
sides as defined by the Schwarzschild, MONDian, and quantum $r(m)$ functions of Eqs (\ref{rS},\ref{rM},\ref{Compton})},
with its lower side (\ref{Compton}) defining the quantum boundary,
specifies the region where classical Newtonian physics is supposed to hold. 
}
\\

The four constants $G$, $h$, $c$, and $a_M$ are of two different types. Indeed, as noted by Milgrom \cite{Milgrom2019},
$G$ is essentially a conversion factor  that links the concepts of gravitational and inertial
masses (akin to the Boltzmann constant $k$ which converts temperature to energy). On the other hand, in line with Milgrom views,
$h$, $c$, and $a_M$ may be considered as physically fundamental concepts somewhat similar to one another: they set the boundaries (constitute the delimiters) of physical applicability regime
corresponding to the classical Newtonian dynamics
(specified by
action, acceleration, and inverse velocity much greater than $h$, $a_M$, and $1/c$).
Thus, (\ref{MCM}) and (\ref{rCM}) seem more physical than (\ref{MCS}) and (\ref{rCS}),
the latter being defined with the help of only two (out of the available three) `limiting' fundamental physical constants.\\

{The coincidence of $r_{CM}$ and $m_{CM}$ with the hadronic scales $r_H$ and $m_H$ was noted already by Milgrom \cite{Milgrom1983}. It was later discussed in \cite{Capozziello2011} by Capozziello and coworkers, who argued
that $a_M$ indeed plays the role of a new fundamental physical parameter. Their argument was based on the discussion of scaling laws
 connecting various self-gravitating astrophysical systems\footnote 
 {ranging from stars through globular 
 clusters and galaxy superclusters to the whole observed universe and considered as states composed of a very large number of smaller constituents, down to protons themselves} 
 and indicating the emergence
 of  $a_M$ as a characteristic scale ruling such systems (see also Fig. 1 in \cite{Hernandez}). The argument involved the discussion of the generalization of the Dirac mass quantization condition (\ref{MCS}) to actions many times larger than $h$ and the use of the observed scaling laws (see eg. \cite{MPLA15}). It is on the basis of such or similar considerations that
the fundamental nature of the MOND parameter $a_M$ (used by us as an assumption) may be accepted. 
}\\

{
For  the gravitational field induced by mass $m$
the introduction of $a_M$ as a new fundamental parameter in addition to $G$ and $c$ leads
to the emergence of a second "fundamental length" (or gravitational radius). Thus,  in addition to the familiar Schwarzschild radius (\ref{rS}), the MONDian radius (\ref{rM}) appears. The absence of such scale in Newtonian gravity indicates that
 going beyond its Einsteinian relativistic generalization is required. In this connection
it may be noted that $r_M$ may be naturally included in an extended metric theory \cite{EPJC71}.\footnote{In such an approach, Noether's symmetries lead to conserved quantities related
to both relevant length scales.}
}\\

One should further note an important difference in the dependence of ($m_{CM}$, $r_{CM}$) and ($m_P$, $l_P$)
on the fundamental constants $h$ and $c$ involved in the determination of these two 
sets of scales.
Indeed, while the set ($m_P$, $l_P$) depends both on $hc$ and $h/c$, the set ($m_{CM}$, $r_{CM}$)
depends on $h/c$ only.
Within the scheme of \cite{ZenMPLA2019} this latter set is obtained by requesting the absence of $hc$
from formula 
\begin{equation}
m_{\delta}=\frac{ha_M}{c^3}\left(\frac{c^7}{Gha^2_M}\right)^{\delta},
\end{equation}
(Eq. 8 in \cite{ZenMPLA2019}), which is achieved for $\delta=1/3$.
Furthermore,
there is an important conceptual difference between the ways the characteristic 
mass and distance scales were estimated in
\cite{ZenMPLA2019} and in Eqs (\ref{MCM},\ref{rCM}) above. Namely, 
the condition used in \cite{ZenMPLA2019} (see also \cite{Burikham}) was that of mathematical simplicity, i.e. the independence of the characteristic scales of one of three constants ($h$, $c$, and the cosmological constant $\Lambda$). On the other hand, the
estimate of Eqs (\ref{MCM},\ref{rCM}) is based on a physical condition: the identification of  mass and distance appearing in the MOND and in the quantum mass-distance relations (\ref{CM}).
Thus, when compared to \cite{ZenMPLA2019}, the derivation leading  to hadronic scales seems to be more physical than the derivations leading to Planck scales and other scales discussed in \cite{ZenMPLA2019}.
\\

In spite of the quantum relation (\ref{Compton}) connecting mass $m_{CM}$ and distance $r_{CM}$,  there exists a different link between the two scales, obtained by elimination of $h/c$ from (\ref{MCM},\ref{rCM}) and thus
independent of the quantum constant $h$ (this follows from Eq. (\ref{rM}) being accepted in the region of the `small' as well):
\begin{equation}
\label{mCMrCM}
m_{CM}=r^2_{CM}\frac{a_M}{G}.
\end{equation}
By Eq. (\ref{mCMrCM}) the region of `small masses' is connected  to the region of `small distances',
a link satisfying the condition expected in \cite{Zen2018} to be relevant for the transition from the classical
(the `large') to the quantum (the `small') world.
Indeed, in \cite{Zen2018} it was argued that the masses and distances relevant for `quantum gravity' should be both small when compared to the scales of the classical macroscopic world, a condition not satisfied by the set ($m_P$, $l_P$) as Planck mass is of the order of the mass of a flea.\\

As observed by Milgrom \cite{Milgrom1983} the coincidence of $m_{CM}$ and $r_{CM}$ with the hadronic scales is related to the celebrated Eddington-Weinberg-Zeldovich \cite{EW,Zeldovich} formula that may be  written in the form
\begin{equation}
\label{EWZ}
\left(\frac{h^2}{G}\sqrt{\frac{\Lambda}{3}} \right)^{1/3} = 0.34 \times 10^{-24}g \approx m_H,
\end{equation}
where $\Lambda= 1.19 \times 10^{-56} cm^{-2}$
is the cosmological
constant and $m_H$ is of the order of pion or nucleon mass \cite{Zen2018}. The connection between (\ref{MCM}) and (\ref{EWZ}) (and thus the cosmological constant argument of \cite{Zen2018})
follows from
`cosmic coincidences' between various astrophysical/cosmological acceleration parameters (which are not understood as yet, see eg. Milgrom's opinion in \cite{Milgrom2020}) 
\begin{equation}
\label{cosmicLambda}
cH_0 \approx c^2\sqrt{\Lambda}\approx 8.2~ a_M.
\end{equation}
{\color{black} where $H_0$ is the present day value of the variable expansion rate of the Universe.}\\

There are two basic ways of ensuring a modification of Newtonian dynamics as given in Eq. (\ref{MOND1}): either through
the modification of the law of inertia or through the modification of the Newtonian force
of gravity. We are inclined to accept the second option as it seems to be more in line with the above-discussed conceptual ideas on the role of gravitational field for the emergence of space.\footnote{This option is also more popular in the attempts
to put the phenomenological idea of MOND on a deeper theoretical basis.} 
Two further aspects of formula (\ref{MOND1}) are then of interest. The first is the proportionality
 of $a$ to $1/r$. As
the proportionality of Newtonian force to $1/r^2$ is associated with three dimensions of ordinary space,
the proportionality of $a$ to $1/r$
suggests (in my opinion) that in the deep-MOND region the lines of gravitational field are restricted/squeezed  
to two dimensions only. 
This may be a hint
that the familiar {matter-and-field-induced} 3D space is somehow constructed from its 2D subspaces, an observation that may be relevant for the idea of space emergence.
{In this connection the important question seems to be: how weak gravitational (MONDian) fields
combine to form the stronger fields of the Newtonian description.}\\

The second feature of formula (\ref{MOND1}) is the proportionality of $a$ to $\sqrt{m}$ 
(in place of the Newtonian proportionality to $m$)
which looks very strange.
It looks even more
peculiar 
when viewed in connection with the MOND-induced quantum mass scale (\ref{MCM})
which fits well the hadronic/elementary particle mass scale (except for neutrinos).
Namely, it turns out 
that the spectrum of elementary particles - at least in the case of charged leptons - does indicate
 a possibly fundamental role for the square root of mass. 
 Specifically,  the physical (ie. the pole but {\it not} the running) masses of charged leptons $m_l$ 
($l=e,\mu, \tau$)
appear to satisfy the Koide relation \cite{Koide}
\begin{equation}
\frac{m_e+m_{\mu}+m_{\tau}}{(\sqrt{m_e}+\sqrt{m_{\mu}}+\sqrt{m_{\tau}})^2}= \frac{2}{3},
\end{equation}
with the left-hand side measured to be $0.666661(7)$. 
The fact that a simple function of three seemingly unrelated 
(square roots of) lepton masses is --
within experimental errors -- exactly
equal to a ratio of two small integers is highly puzzling.
Although the original Koide formula refers to leptons only, related regularities have been observed for quarks as well (see eg. \cite{ZenKoide}).
Together, these observations seem to indicate the fundamental
relevance of the square root of mass in the spectrum of elementary fermions. 
{
The relevance of the particle/hadronic mass scale for the idea of space emergence seems 
to be further boosted by baryon phenomenology.
Indeed, as stressed in \cite{Capstick}, the observed spectrum of excited baryons
(ie. the absence of many, many baryonic states predicted in standard quark approaches) 
strongly suggests that one of the two internal spatial degrees of (quark) freedom in excited baryons is completely frozen.  
}\\

The above MOND-based considerations 
show an essential difference between the ways Planck and hadronic mass-and-distance scales depend on fundamental constants $h$ and $c$ (as well as $a_M$), point out the possible relevance of 2D subspaces of the 3D space, 
and stress the role played by the square root of mass
(both in the gravitational and
in the particle settings).
Thus, they  
seem to provide further argument for the idea advocated in \cite{Zen2018} that the spectrum of elementary particles holds important information
concerning the nature of space.

\vfill

\vfill

\end{document}